\documentclass[journal]{IEEEtran}

\usepackage[utf8]{inputenc}
\usepackage{multirow}
\usepackage{comment}
\usepackage{xspace}
\usepackage{balance}
\usepackage{multirow}
\usepackage[table,xcdraw]{xcolor}
\usepackage{mathtools}
\usepackage{booktabs}
\usepackage{amssymb}
\usepackage{pdfpages}
\usepackage{graphicx}
\usepackage{caption}
\usepackage{subcaption}
\usepackage{caption, subcaption}
\usepackage{amsthm}
\usepackage{float}
\usepackage{url}
\usepackage{cleveref}
\usepackage{flushend}
\usepackage[absolute,overlay]{textpos}

\newcommand\blfootnote[1]{%
  \begingroup
  \renewcommand\thefootnote{}\footnote{#1}%
  \addtocounter{footnote}{-1}%
  \endgroup
}

\newcommand{\httparchive}{{\emph{HTTPArchive} \xspace}}
\newcommand{\browsertime}{{\emph{BrowserTime} \xspace}}

\title{Measuring HTTP/3:\\Adoption and Performance}

\author{
Martino Trevisan$^\dagger$,
Danilo Giordano$^\dagger$,
Idilio Drago $^\ddagger$,
Ali Safari Khatouni$^\star$\\
\small{$^\dagger$Politecnico di Torino,
$^\ddagger$University of Turin,
$^\star$Shopify} \\
\texttt{first.last@\{polito.it, unito.it, shopify.com\}}
}

\begin{document}

\maketitle

\TPshowboxestrue
\TPMargin{0.15cm}
\begin{textblock*}{17.5cm}(2cm,0.3cm)
\normalsize
\bf
\definecolor{myRed}{rgb}{0.55,0,0}
\color{myRed}
\noindent
Please cite this article as: Martino Trevisan, Danilo Giordano, Ali Safari Khatouni. Measuring HTTP/3: Adoption and Performance. 19th Mediterranean Communication and Computer Networking Conference (2021). DOI: \url{https://doi.org/10.1109/MedComNet52149.2021.9501274}
\end{textblock*}

\begin{abstract}

The third version of the Hypertext Transfer Protocol (HTTP) is in its final standardization phase by the IETF. Besides better security and increased flexibility, it promises benefits in terms of performance. HTTP/3 adopts a more efficient header compression schema and replaces TCP with QUIC, a transport protocol carried over UDP, originally proposed by Google and currently under standardization too. Although HTTP/3 early implementations already exist and some websites announce its support, it has been subject to few studies.
We provide a first measurement study on HTTP/3 adoption and performance. We testify how it has been adopted by some of the leading Internet companies such as Google, Facebook and Cloudflare in 2020. We run a large-scale measurement campaign towards thousands of websites adopting HTTP/3, aiming at understanding to what extent it achieves better performance than HTTP/2. We find that adopting websites often host most web page objects on third-party servers, which support only HTTP/2 or even HTTP/1.1. As excepted, websites loading objects from a limited set of third-party domains (avoiding legacy protocols) are those experiencing larger performance gains. Our experiments however show that HTTP/3 provides sizable benefits only in scenarios with high latency or poor bandwidth. \blfootnote{
This work has been supported by the EU H2020 research and innovation programme under grant agreement No. 644399 (MONROE) and by the SmartData@PoliTO center on Big Data and Data Science.}
\end{abstract}


\section{Introduction}
\label{sec:introduction}

The Hypertext Transfer Protocol (HTTP) is the king of web protocols and is used to access the vast majority of services on the Internet, from websites to social networks and collaborative platforms. HTTP was born in the early 90s, and its first version 1.1 was standardized in 1997~\cite{rfc2068}. Only in 2014, HTTP/2~\cite{rfc7540} was proposed, with substantial changes in the framing mechanisms. HTTP/3 is the third version of HTTP and is currently in the final standardization phase at the IETF~\cite{ietf-quic-http-33}. It promises performance benefits and security improvements compared to HTTP/2. As a major change, HTTP/3 replaces TCP as transport layer in favor of QUIC, a UDP-based protocol originally proposed by Google and currently being standardized by the IETF~\cite{ietf-quic-transport-33}. Furthermore, it introduces a more effective header compression mechanism and exploits TLS 1.3~\cite{rfc8446} (or higher) to improve the level of security.

HTTP/3 is expected to take over the place of HTTP/2 in the next years, and some of the leading Internet companies already announced its support during 2020 (e.g., the CloudFlare CDN\footnote{\url{https://blog.cloudflare.com/http3-the-past-present-and-future/}} and Facebook\footnote{\url{https://engineering.fb.com/2020/10/21/networking-traffic/how-facebook-is-bringing-quic-to-billions/}}). However, currently neither the real state of its deployment nor the performance benefits of HTTP/3 have been measured yet. 

In this paper we fill this gap by running the first large-scale measurement study on HTTP/3 adoption and performance. We first rely on the HTTPArchive Dataset to study to what extent the web ecosystem has already adopted HTTP/3. Then, we run additional campaigns to measure the benefits introduced by HTTP/3. Considering websites that adopt different versions of the HTTP protocol, we measure several metrics known to indicate users' Quality of Experience (QoE). Finally, we emulate different network conditions on the paths connecting our measurement platform to assess whether and how HTTP/3 improves performance in different scenarios. 

Using the open-source HTTPArchive Dataset,\footnote{https://httparchive.org/} we find thousands of websites supporting HTTP/3, most of them hosted by a handful of Internet hyper-giants, i.e., Facebook, Google, and Cloudflare. We then automatically revisit websites supporting HTTP/3 under diverse network conditions to measure the performance benefits in terms of QoE-related metrics. We visit $14\,707$ websites in total while emulating artificial latency, packet loss, and limiting the bandwidth. We run $2\,647\,260$ visits over a period of one month. We find that HTTP/3 benefits emerge only on particular network conditions and strongly differ across websites. Our key findings are:
\begin{itemize}
    \item Google, Facebook and Cloudflare are the early adopters of HTTP/3, hosting almost the totality of currently websites supporting HTTP/3.
    \item The majority of web page objects in websites supporting HTTP/3 are still hosted on non-HTTP/3 third-party servers.
    \item We observe sizable performance benefits only in scenarios with high latency or low network bandwidth.
    \item The performance gains largely depends on the infrastructure hosting the website, possibly due to optimizations on server-side infrastructure.
    \item As expected, the websites relying on fewer connections to load objects are those benefiting the most.
\end{itemize}


The remainder of the paper is organized as follows: Section~\ref{sec:background} describes HTTP/3 and illustrates related work. Section~\ref{sec:dataset} presents our datasets and how we have collected them. Section~\ref{sec:results} illustrates our results on HTTP/3 adoption and performance. Section~\ref{sec:limitations} discusses limitations of results and lists the basis for future work, while Section~\ref{sec:conclusion} concludes the paper.

\section{Background}
\label{sec:background}

\subsection{HTTP/3}
\label{sec:h3}

\begin{figure}[t]
    \centering
    \includegraphics[width=0.8\columnwidth]{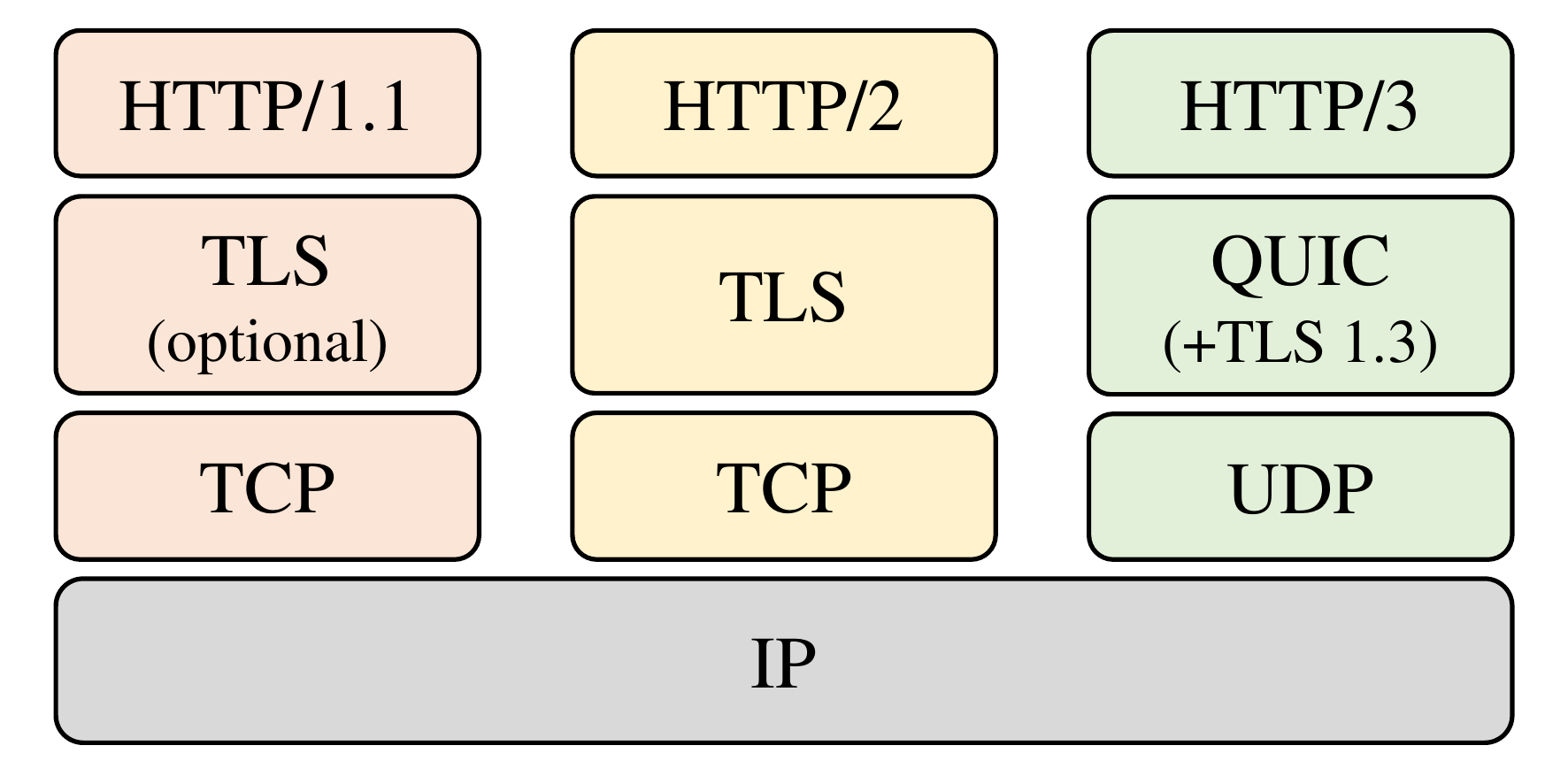}
    \caption{Protocol stack for different HTTP versions.}
    \label{fig:protocol_stack}
\end{figure}

HTTP/3 is the third version of the well-known Hypertext Transfer Protocol, born in the 90s to transfer multimedia content and hyper-textual documents over the Internet. With its version 1.1, it has been the king of web protocols for more than 20 years, superseded only by its second version HTTP/2 in 2014. HTTP/2 implemented several novel features, especially to improve how data is framed and transported. It promised to make the web faster, even if some studies questioned its benefits~\cite{varvello2016web,rajiullah2019web}.

HTTP/3 is currently in the final standardization phase, reaching the 34$^{th}$ draft version~\cite{ietf-quic-http-33} and making it stable and usable for real deployments. The main improvements from version 2 include more efficient header compression, advanced security features based on TLS 1.3, and, especially, the use of QUIC at the transport layer. The resulting protocol stack is thus heavily modified, as we show in Figure~\ref{fig:protocol_stack}. QUIC, initially developed by Google, is a transport protocol based on UDP and is currently in the standardization phase too~\cite{ietf-quic-transport-33}. QUIC revisits TCP, moving congestion control in user space and allowing faster handshakes. Moreover, it solves the long-standing issue of head-of-line blocking, allowing multiple independent streams within the same connection. Indeed, QUIC allows independent retransmission for sub-streams and decouples it from congestion control. This operation is expected to improve users' QoE with faster website responsiveness, especially in scenarios with poor network conditions. HTTP/3 also mandates the use TLS 1.3~\cite{rfc8446}, directly incorporated at the QUIC layer. Finally, it allows 1-RTT handshakes and 0-RTT resumption, further reducing session setup time.

\subsection{Related Work}
\label{sec:related}

Given its recent conception, few works already targeted HTTP/3.
Saif~\emph{et al.}~\cite{saif2020early} run experiments controlling both client and server accessing a single web page. They study the effect of delay, packet loss and throughput, without finding any major impact on performance. In contrast to them, we run a large-scale measurement campaign, controlling only the client and targeting thousands of HTTP/3 websites residing on their original servers.
Marx~\emph{et al.}~\cite{marx2020same} compare $15$ HTTP/3 implementations, finding a large heterogeneity in how congestion control, prioritization and packetization work. They only run single file downloads, but their results call for extensive in-the-wild measurements.
Cloudflare benchmarks its own draft 27 HTTP/3 implementation in~\cite{cloudflareh3}, finding it to be $1-4$\% slower than HTTP/2. However, their experiments are limited to the \texttt{blog.cloudflare.com} website.
Guillen~\emph{et al.}~\cite{guillen2019sand} proposed a control algorithm for adaptive streaming tailored for HTTP/3.

Google proposed QUIC in 2012 and, as such, it has been the subject of many studies. Wolsing~\emph{et al.}~\cite{wolsing2019performance} show that QUIC outperforms TCP thanks to the fast connection setup.
Manzoor~\emph{et al.}~\cite{manzoor2020performance} show that QUIC performs worse than TCP in Wireless Mesh Networks thanks to bad interactions of the protocol with the WiFi layer in that scenario.
Carlucci~\emph{et al.}~\cite{carlucci2015quic} found that QUIC reduces the overall page retrieval time. Kakhi~\emph{et al.}~\cite{kakhki2017taking} run a large-scale measurement campaign on QUIC, finding that it outperforms TCP in most cases. These works however target Google's QUIC versions, while the current standard proposed at the IETF has made significant progresses~\cite{kosek2021beyond}. Moreover, they focus uniquely on transport layer, neglecting the improvement introduced by HTTP/3 in higher layers, which we measure in this work.

\section{Data Collection}
\label{sec:dataset}

We rely on two datasets to study (i) the adoption of HTTP/3 and (ii) its performance on diverse network conditions. We summarize them in Table~\ref{tab:datasets}.

\begin{table}[t]
    \centering
    \normalsize
    \def\arraystretch{1.2}
    \caption{Description of the employed datasets.}
    \label{tab:datasets}
    \begin{tabular}{lll}
        \textbf{Dataset} & \textbf{Visits} & \textbf{Goal} \\\hline
        \httparchive & $53\,107\,185$ & HTTP/3 Adoption\\
        \browsertime & $2\,647\,260$ & HTTP/3 Performance\\\hline
    \end{tabular}
\end{table}

\subsection{HTTP/3 Adoption -- HTTPArchive}

We study the adoption of HTTP/3 using the HTTPArchive, an open dataset available online.\footnote{\url{https://httparchive.org/}, visited on February 4th 2021.} The dataset contains metadata coming from visits to a list of more than $5$ million URLs provided by the Chrome User Experience Report.\footnote{\url{https://developers.google.com/web/tools/chrome-user-experience-report}} The list of URL is built using the navigation data of real Chrome users and contains a representative view of the most popular website and web services accessed worldwide.\footnote{HTTPArchive used to adopt the Alexa top-1M website list, but switched to the Chrome User Experience Report when Alexa discontinued the rank in July 2018.} Each month, all URLs are visited using the Google Chrome browser from a U.S. data center, and the resulting navigation data is made public. For each visit, the dataset contains information about the page characteristics, loading performance, as well as the HTTP transactions in HAR format,\footnote{\url{http://www.softwareishard.com/blog/har-12-spec/}} including request and response headers.

Fundamental for our analyses, the details of HTTP responses indicate the eventual \texttt{Alt-Svc} header, which is used by servers to announce support to HTTP/3. By setting the \texttt{Alt-Svc} header, the server has the possibility to inform the client to make subsequent connections using HTTP/3 and may specify the support to specific draft versions (e.g., $27$ or $29$).

We download the HTTPArchive dataset starting from November $2019$, when we observe the first websites offering support to HTTP/3. We use the data to study the trend of HTTP/3 adoption. The data sum up to $6.6$ TB. Since we are interested in studying the adoption of HTTP/3 on \emph{websites}, we discard all visits to internal pages (less than half of the total) and keep only visits to home pages. We refer to this dataset as \emph{HTTPArchive}.

\subsection{HTTP/3 Performance -- BrowserTime}

We use the most recent snapshot at the time of writing (December 2020) to build the list of websites currently supporting HTTP/3. We find $14\,707$ websites announcing support to HTTP/3. Next, we visit these websites with three HTTP versions (HTTP/1.1, HTTP/2, and HTTP/3) to quantify possible performance improvements. To this end, we rely on BrowserTime, a dockerized tool to run automatic visits to web pages with a large set of configurable parameters.\footnote{\url{https://www.sitespeed.io/documentation/browsertime/}} We use BrowserTime to instrument Google Chrome to visit websites using a specific HTTP version. Important for our goal, Google Chrome allows specifying a set of domains to be contacted with HTTP/3 on the first visit, i.e., without prior indication via \texttt{Alt-Svc} header. We limit ourselves to Chrome, since we are not aware of similar functionalities in other browsers (e.g., Firefox).

\begin{table}[t]
    \centering
    \normalsize
    \def\arraystretch{1.2}
    \caption{Network configurations used in the experiments.}
    \label{tab:impairment}
    \begin{tabular}{ll}
        \textbf{Parameter} & \textbf{Tested configurations} \\\hline
        Latency [ms]& Native, 50, 100, 200\\
        Loss [\%]& Native, 1, 2, 5\\
        Bandwidth [Mbit/s] & Native, 5, 2, 1\\ \hline
    \end{tabular}
\end{table}

We are interested in studying the impact of HTTP/3 under different network conditions. As such, we run our measurements enforcing different \textit{network configurations}. We run our experiments using two high-end servers connected to the Internet via $1$ Gbit/s Ethernet and located in our university campus. We call this baseline scenario \emph{Native}, as reported in Table~\ref{tab:impairment}.

Then, we enforce other configurations during the visits relying on the well-known Linux \texttt{tc} tool. Each network configuration is defined by changing one of three parameters: (i) extra latency, (ii) extra packet loss, or (iii) bandwidth limit. For each parameter, we use $4$ different configurations, reported in Table~\ref{tab:impairment}. In case of latency, we impose it on the uplink, while packet loss and bandwidth limit are enforced on both up and down links. For each network configuration, we visit each website (i) enabling only HTTP/1.1, (ii) enabling HTTP/1.1 and HTTP/2, and (iii) enabling all three versions of the protocol. All visits to the same website are run consecutively, cleaning all state between repetitions, i.e., browser cache, TCP connections etc. Visits are repeated $5$ times to get more reliable results. Hence, we visit each website $4 \times 3 \times 3 \times 5 = 180$ times. 

BrowserTime collects several statistics for each visit, including details on all HTTP transactions as well as performance metrics. We track two metrics that have been shown to be correlated with users' QoE~\cite{da2018narrowing} and can be estimated also at the ISPs~\cite{trevisan2019pain}:

\begin{itemize}
    \item \textbf{onLoad}: The time at which the browser fires the \texttt{onLoad} event -- i.e., when all elements of the page, including images, style sheets and scripts have been downloaded and parsed;
    \item \textbf{SpeedIndex}: Proposed by Google,\footnote{\url{https://web.dev/speed-index/}} it represents the time at which visible portions of the page are displayed. It is computed by capturing the video of the browser screen and tracking the visual progress of the page during rendering.
\end{itemize}

In total, we run $2\,647\,260$ visits over a period of one month. The visit metadata account for $189$ GB, and we call this dataset \emph{BrowserTime}.

\section{HTTP/3 adoption and performance}
\label{sec:results}

In this section we first provide an overview of the HTTP/3 adoption (Section~\ref{sec:adoption}). Since announcing HTTP/3 support is not the same as serving content using the protocol, we quantify the amount of content served over HTTP/3 (Section~\ref{sec:content}). Then, we study how HTTP/3 affects QoE-related performance metrics  (Section~\ref{sec:controlled}) and whether identified improvements can be related to the provider hosting content (Section~\ref{sec:provider}) or website characteristics (Section~\ref{sec:website}).

\subsection{HTTP/3 adoption}
\label{sec:adoption}

\begin{figure}[t]
    \centering
    \includegraphics[width=\columnwidth]{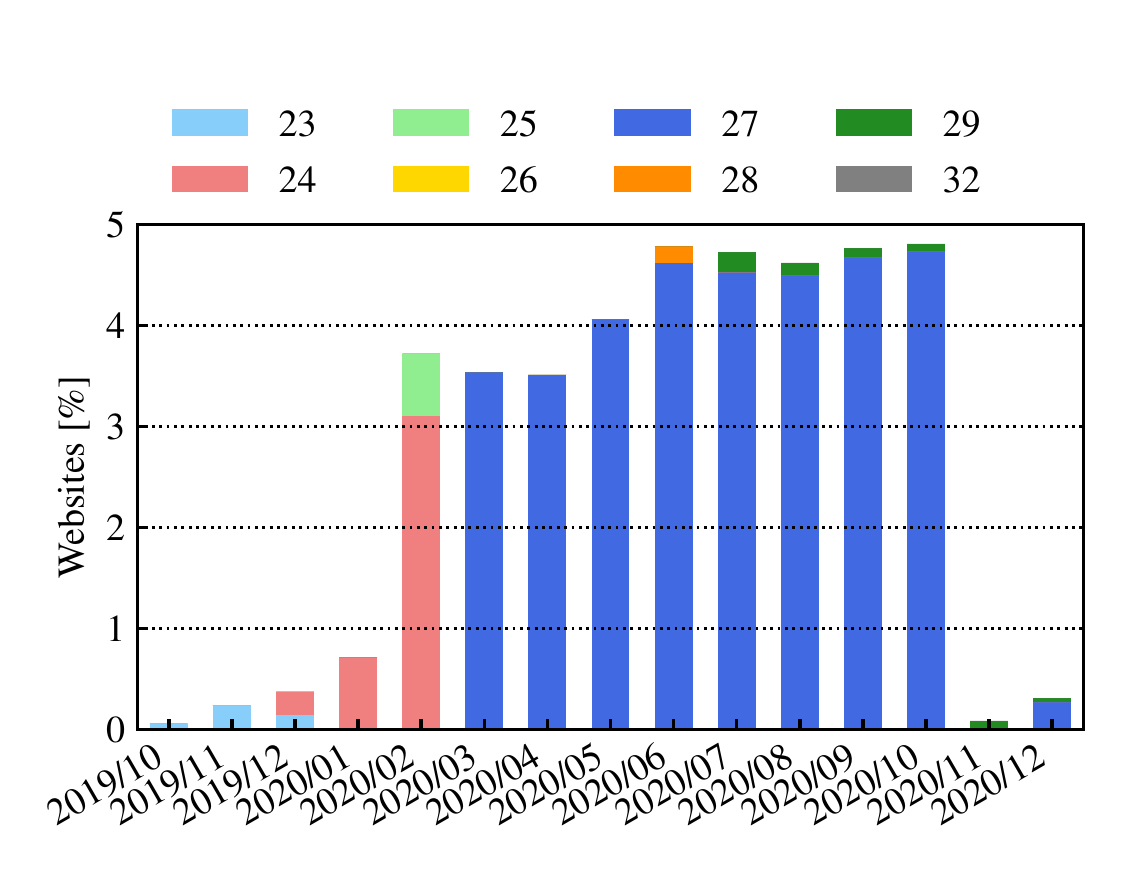}
    \caption{Percentage of websites in HTTPArchive that announce support to HTTP/3, separately by IETF draft.}
    \label{fig:versions_time}
\end{figure}

\begin{figure}[t]
    \centering
    \includegraphics[width=\columnwidth]{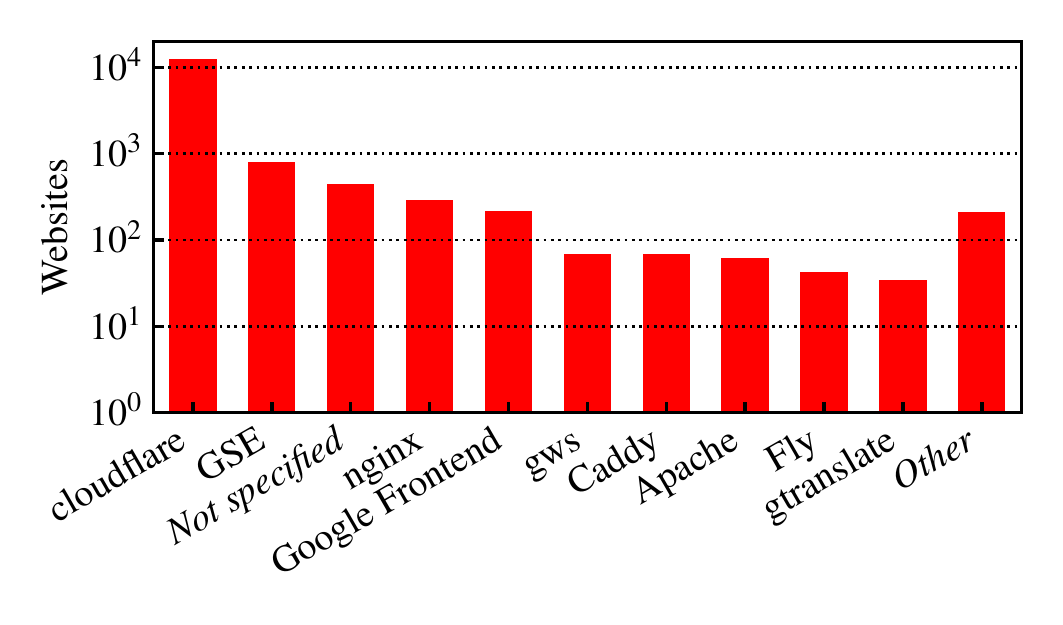}
    \caption{\texttt{Server} in HTTP response (December 2020).}
    \label{fig:server}
\end{figure}

\begin{figure}[t]
    \centering
    \includegraphics[width=\columnwidth]{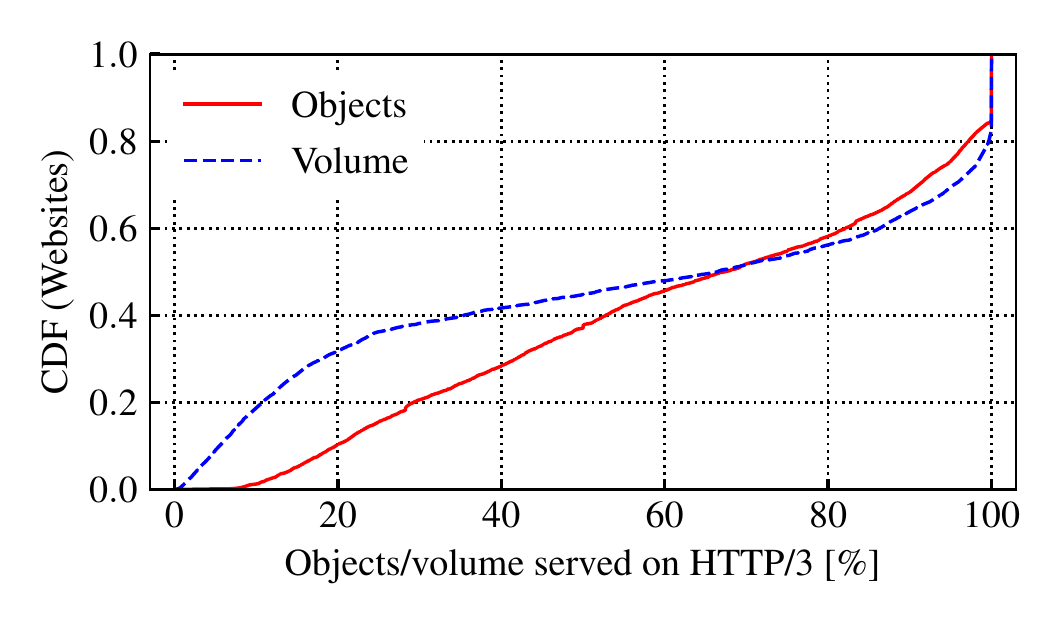}
    \caption{Share of objects/volume served using HTTP/3 on enabled websites.}
    \label{fig:share_objects}
\end{figure}

We first study to what extent HTTP/3 has been adopted since its first proposal. To this end, we profit from the \httparchive dataset. The first IETF draft was published on in January 2017, but we observe the first websites adopting HTTP/3 only in late 2019. Since then, the number of websites supporting HTTP/3 has started to grow. Figure~\ref{fig:versions_time} shows the trend for the last months of 2019 and the entire 2020. Looking at the \texttt{Alt-Svc} header, we can observe the HTTP/3 draft version supported by the server, shown with different colors in the figure. In case a website offers more than one version, we considered the earliest observed in \emph{HTTPArchive}. 

In the first four months, the number of websites supporting HTTP/3 increased slowly, reaching 0.7 \% of the total. At that period, only Google and Facebook used to offer HTTP/3 for their websites. In February 2020, the number of websites supporting HTTP/3 exploded. This is due to CloudFlare, which enabled HTTP/3 on most of the websites it hosts. The share of websites supporting HTTP/3 passes $4$\%, reaching a maximum in October 2020, with $4.8$\% of the websites ($203$~k). In November, the number of websites suddenly dropped to less than $0.1$ \% ($4\,024$ in absolute terms). This was caused by CloudFlare suspending support to HTTP/3 due to performance issues, as declared online.\footnote{\url{https://community.cloudflare.com/t/community-tip-http-3-with-quic/117551}, visited on 2/20/2021.} On December, CloudFlare re-enabled HTTP/3 on a subset of websites. On that date, $14\,707$ websites were announcing support to HTTP/3. Since we need to revisit websites to measure their performance, we consider only these $14\,707$ websites for results that will follow.

The majority of the $14\,707$ websites supporting HTTP/3 are hosted by large companies running their own server applications. We breakdown these numbers in Figure~\ref{fig:server}, which indicates the $10$ most popular servers as indicated on the HTTP \texttt{Server} header. CloudFlare, as expected, hosts most of the websites supporting HTTP/3 (notice the log $y$-scale). Google is in the second position, with GSE (Google Servlet Engine), Google Frontend and GWS (Google Web Server). Indeed, GSE is used on the Blogspot platform, represented by $809$ websites in our list. For $575$ websites, there is no \texttt{server} indication on HTTP responses, and we find that $445$ of these websites belong to Facebook -- e.g., \texttt{facebook.com} and \texttt{instagram.com} domains. The remaining websites run popular open-source servers (nginx, Apache) or more peculiar HTTP ones (e.g., Caddy) that offer HTTP/3 support in their earlier versions. 

\subsection{Content served over HTTP/3}
\label{sec:content}

Next, we study to what extent objects of enabled websites are served using HTTP/3. Indeed, even if a website supports HTTP/3, not \emph{all} of its objects are served through HTTP/3. Objects may be downloaded from external CDNs, cloud providers or third-parties not supporting the same protocol. This is the case, for example, for ads and trackers typically hosted on different third-party infrastructure. We use the \browsertime dataset, which allows us to observe the protocol used for delivering each object composing the visited websites. 

In Figure~\ref{fig:share_objects} we consider all visits run with HTTP/3 enabled. For each visit, we compute the share of objects served over HTTP/3. As each website is accessed multiple times, we average the values across visits. Clearly, at least the main HTML document is always sent over HTTP/3, but the remaining objects may be served with older HTTP versions. The figure shows the distribution of the percentage of objects served over HTTP/3 (solid red line) and also depicts the byte-wise percentage -- i.e., weighting each object by its size. We first notice that in $18$\% of cases, all objects are delivered over HTTP/3, meaning that the web page only contains elements hosted on HTTP/3 enabled servers. The websites having $90$\% or more of objects (volume) on HTTP/3 are $36$ ($41$)~\% and only $9$ ($28$) \% have less than $20$ \% of objects (volume). Interestingly, we notice that $51$\% of websites still have one or more objects retrieved using HTTP/1.1.

\begin{figure}[t]
    \centering
    \includegraphics[width=\columnwidth]{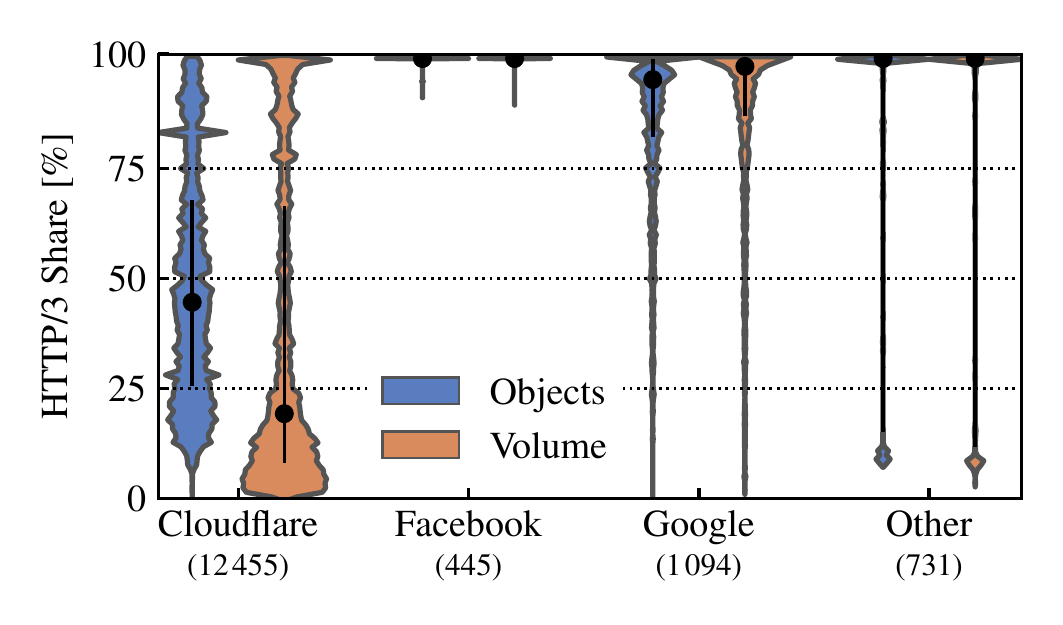}
    \caption{Share of objects/volume served on HTTP/3, separately by provider.}
    \label{fig:share_objects_cdn}
\end{figure}

Next, we dissect the above analysis by provider -- i.e., the company/CDN hosting the website. We obtain it by looking at the \texttt{server} HTTP header, website name and server IP address, which allow easy identification. As discussed for Figure~\ref{fig:server}, we notice that HTTP/3 has been adopted mostly by (i) Cloudflare CDN, (ii) the Facebook and (iii) Google. The remaining $595$ websites (i.e., Other) belong mostly to self-hosted websites running updated versions of the \emph{nginx} web server. 

Figure~\ref{fig:share_objects_cdn} shows the share of objects and volume served over HTTP/3, separately by provider. Websites hosted by Cloudflare tend to be more heterogeneous, with half of the objects retrieved via non-HTTP/3 servers (on median). Moreover, only 24\% of the volume is served by using HTTP/3. This is likely due to the variety of websites relying on the provider: Indeed, Cloudflare offers its hosting service to a very large number of websites. These websites may use complex web pages composed of several third-party objects stored on external servers that do not rely on HTTP/3 yet. Conversely, Facebook and Google show a very different situation. Almost all objects are served with HTTP/3. This is expected, since Facebook and Google use their CDNs mostly to offer their own services. Looking at Google, the long tail of the distribution is due to Blogspot websites, in which the creator may add content from external sources. Finally, considering the Other category, almost all the objects and volumes are served using HTTP/3. These websites tend to be simple, composed of a few objects stored in the same self-hosted servers together with the main HTML document. 


\subsection{Performance gains}
\label{sec:controlled}

We now study the impact of HTTP/3 on web page performance. To this end, we use the \browsertime dataset, in which the $14\,707$ websites have been visited multiple times under different network conditions. Besides computing the performance in the native scenario (i.e., 1~gpbs Ethernet on a campus network), we use \texttt{tc-netem} to enforce extra latency, packet loss and limit bandwidth. We then contrast page QoE-related performance indicators (onLoad and SpeedIndex), (i) showing their absolute value and (ii) computing a metric that we call \emph{H3 Delta}. Given a website and a given network scenario, we obtain the \emph{H3 Delta} as the relative deviation of the metric when using HTTP/3 ($h3$) instead of HTTP/2 ($h2$). As we always run 5 visits for each case, we consider median values. The \emph{H3 Delta} for a website $w$ in scenario $s$ is computed as follows:

\begin{equation*}
\text{H3-Delta}(w,s) = \frac{\text{median}(w,s, h3) - \text{median}(w,s, h2)}{\text{max}(\text{median}(w,s, h3), \text{median}(w,s, h2))}
\end{equation*}

\noindent
By definition, $\text{H3 Delta}(w,s)$ is bound in $[-1,1]$ and it is negative when a website loads faster under HTTP/3, and positive otherwise. We compute the \emph{H3 Delta} for both onLoad and SpeedIndex.

\begin{figure}[t]
    \begin{flushright}
    \includegraphics[width=\columnwidth]{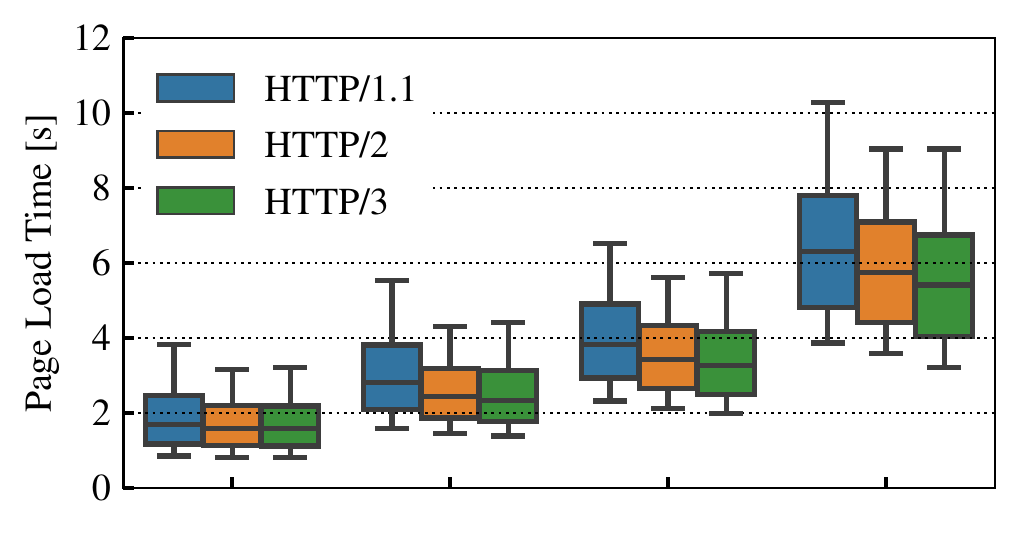}\\
    \includegraphics[width=0.98\columnwidth]{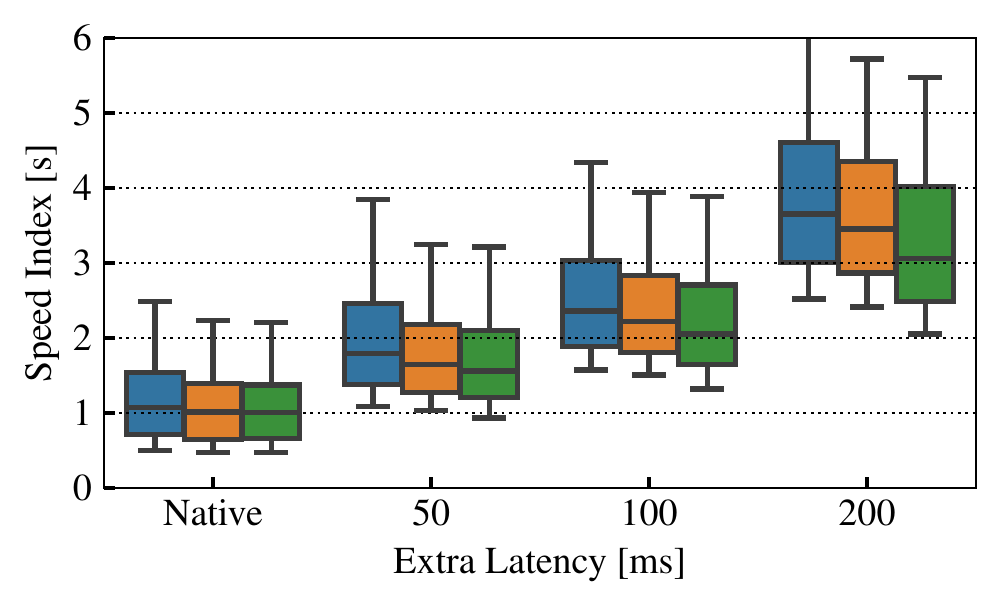}
    \end{flushright}
    \caption{onLoad (top) and SpeedIndex (bottom) with  extra latency, separately  for HTTP/1.1, HTTP/2 and HTTP/3.}
    \label{fig:latency_absolute}
\end{figure}

We illustrate how the metric values vary when imposing different network conditions, focusing firstly on extra latency in Figure~\ref{fig:latency_absolute}. Using boxplots, we show the distribution of onLoad (top) and SpeedIndex (bottom), separately by HTTP version (colored boxes). The boxes span from the first to the third quartile, whiskers report the 10$^{th}$ and the 90$^{th}$ percentiles, while black strokes represent the median. When no extra latency is imposed (\emph{native} case), we observe that onLoad time is in median around $2$s, while SpeedIndex around $1$s, without significant differences across HTTP versions. When adding extra latency, the websites load slower as more time is needed to download the page objects, requiring in median 6 seconds with $200$ ms of additional latency. Not shown here for brevity, also packet loss and limited bandwidth cause similar degradation of performance indicators. Figure~\ref{fig:latency_absolute} shows that HTTP/1.1 has the worst performance with high latency, while HTTP/3 shows the greatest benefits. Considering additional latency of $200$ ms, websites onLoad in median in $6.4$, $5.8$ and $5.4$~s with HTTP versions 1.1, 2 and 3, respectively.

\begin{figure*}[t]
    \centering
    \begin{subfigure}{0.345\textwidth}
        \includegraphics[width=\columnwidth]{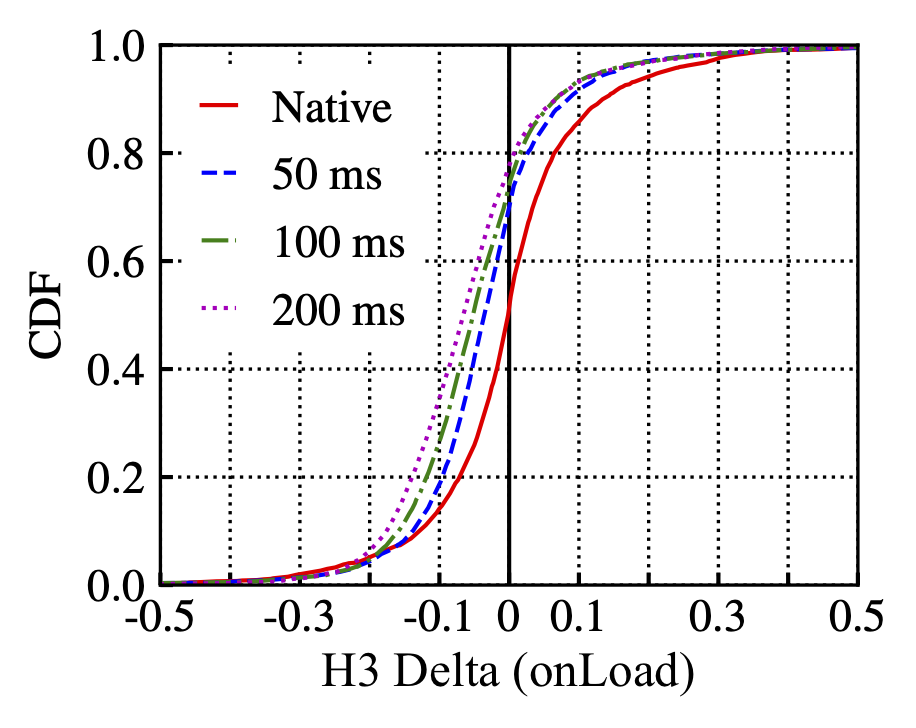}\\
        \includegraphics[width=\columnwidth]{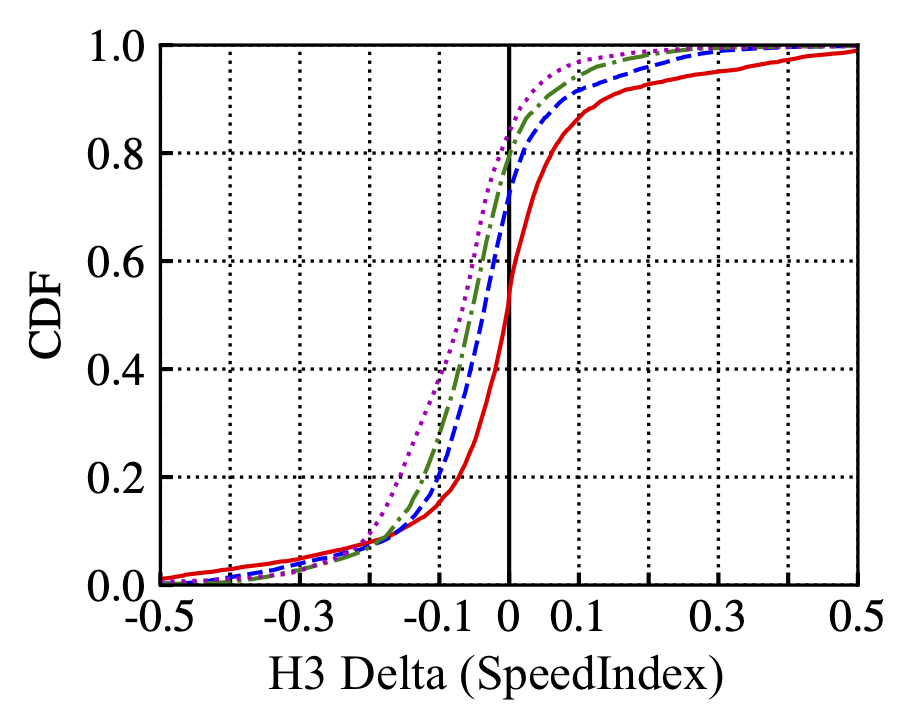}
        \caption{Latency.}
        \label{fig:speedup_latency}
    \end{subfigure}
    \begin{subfigure}{0.31\textwidth}
        \includegraphics[width=\columnwidth]{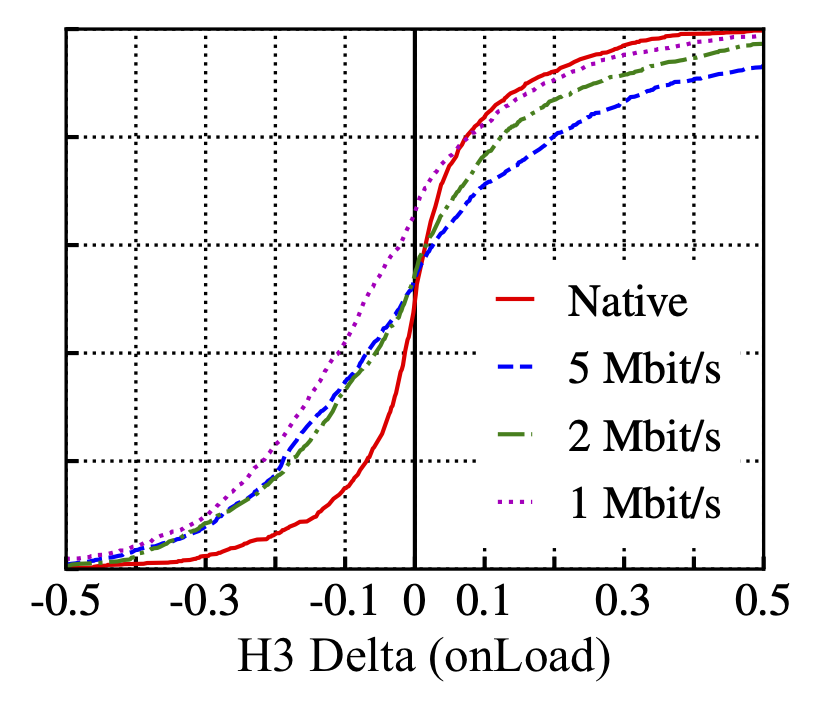}\\
        \includegraphics[width=\columnwidth]{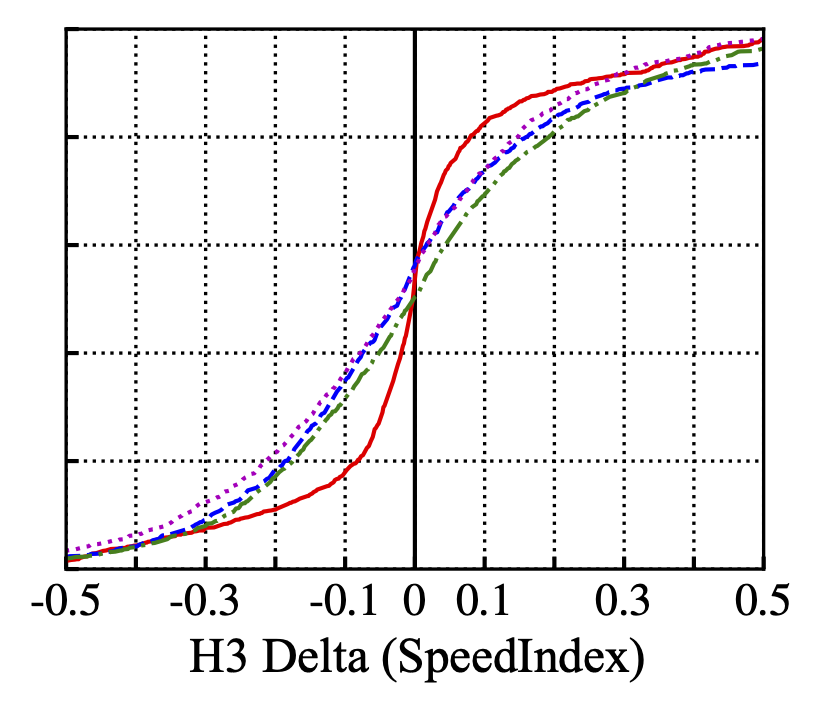}
        \caption{Bandwidth.}
        \label{fig:speedup_bw}
    \end{subfigure}
    \begin{subfigure}{0.31\textwidth}
        \includegraphics[width=\columnwidth]{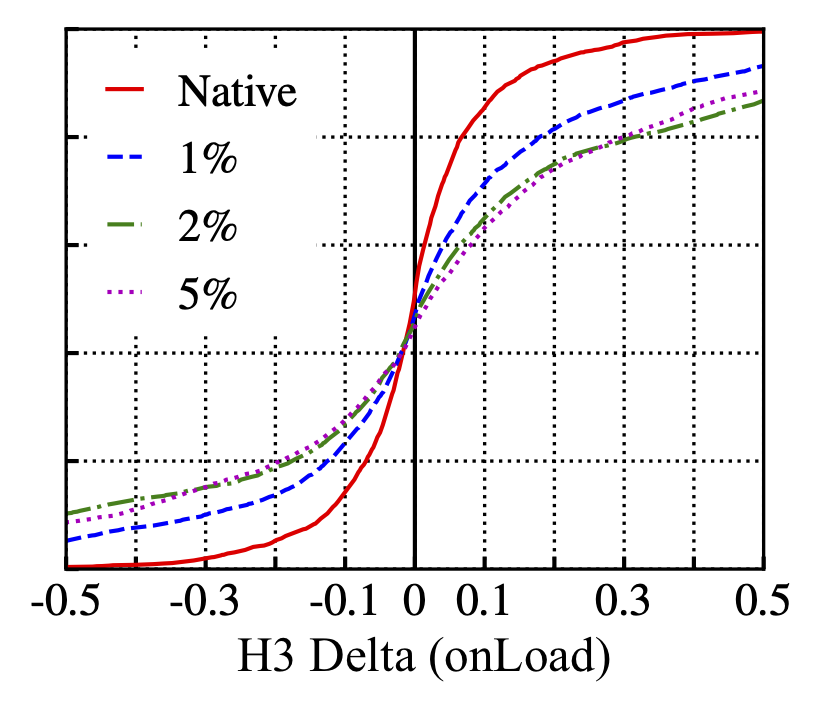}\\
        \includegraphics[width=\columnwidth]{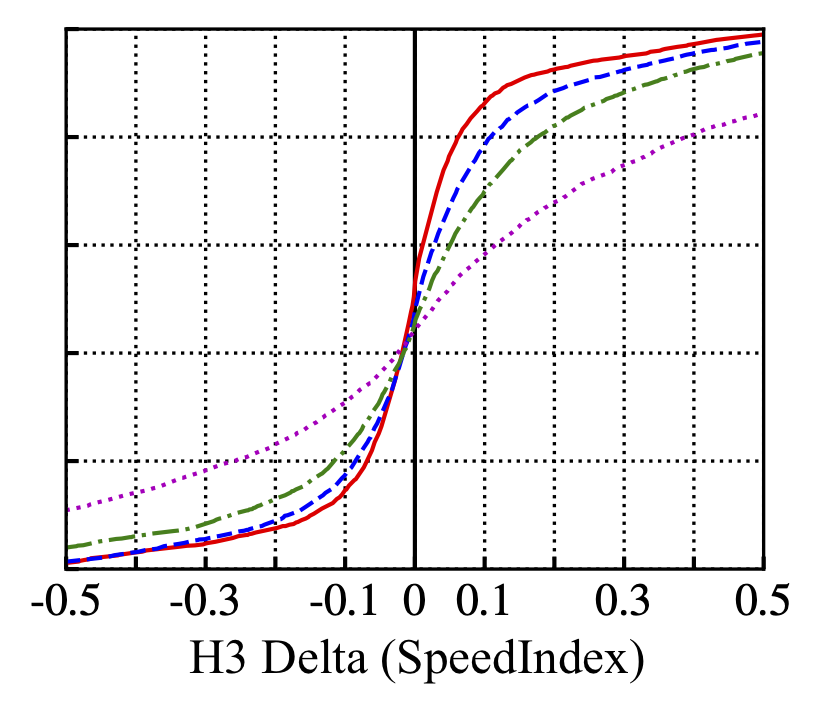}
        \caption{Packet loss.}
        \label{fig:speedup_loss}
    \end{subfigure}
    \caption{\emph{H3 Delta} on different scenarios. onLoad (top) and SpeedIndex (bottom). Negative values indicate that HTTP/3 is faster.}
    \label{fig:speedup}
\end{figure*}

To better catch differences between HTTP/3 and HTTP/2, we now study the \emph{H3 Delta} in Figure~\ref{fig:speedup}, where we show the distribution over the $14\,707$ websites for both onLoad (top row) and SpeedIndex (bottom row). The three columns refer to scenarios with additional latency, limited bandwidth and packet loss, respectively. The solid red lines represent the \emph{native} case. Dashed lines represent scenarios with emulated network conditions, as indicated in Table~\ref{tab:impairment}.

Starting from latency, we confirm what already emerged from Figure~\ref{fig:latency_absolute}. In the native case, we observe no general trend: Looking at the solid red lines, we notice that approximately in 50 \% of the cases websites load faster with HTTP/3 and in the remaining cases HTTP/3 is slower. When latency is high, HTTP/3 gives sizable benefits compared to HTTP/2. If we impose extra latency of $50$ ms, 69 (74) \% of websites have lower onLoad time (SpeedIndex), meaning that they load faster. The number of websites loading faster increase to 76 (81) \% with $100$ ms latency. With $200$ ms, the number of websites loading faster reach 81 (87) \%, and the median \emph{H3 Delta} is -0.08 (0.12). 

Focusing on experiments with bandwidth limitation (central plots in Figure~\ref{fig:speedup}), different considerations hold. We observe sizable benefits only for onLoad time with the bandwith limited to $1$ Mbit/s, where $69$ \% of websites load faster with HTTP/3. Notice that this benefit cannot be introduced by indirect higher latency due to queuing delay (also called bufferbloat), as we limit the machine queues to 32 KB. In other cases, no clear trend emerges, but we notice a larger variability of the \emph{H3 Delta} measure introduced by the constrained setup. For example, in case of SpeedIndex, $56$, $49$, $58$ \% websites load faster with HTTP/3 with $5$, $2$ and $1$ Mbit/s bandwidth, respectively. Similar considerations hold for packet loss (right-most plots in Figure~\ref{fig:speedup}). Despite a larger variability, we cannot identify any general trend, and the \emph{H3 Delta} values are equally distributed above and below $0$. 

In summary, we observe improvements on onLoad time with poor bandwidth when using HTTP/3. HTTP/3 shows sizable benefits in case of high latency. We do not testify performance benefits of HTTP/3 in scenarios with high packet loss and in some other cases. In fact, in several tested cases, some websites can even perform worse when HTTP/3 is enabled. 

%

\subsection{HTTP/3 performance by provider}
\label{sec:provider}

Next we study whether HTTP/3 performance gains could be related to the provider hosting the websites. As we observed sizable performance benefit for HTTP/3 only in cases of high latency or poor bandwidth, we restrict our analyses to those cases.

\begin{figure}[t]
    \vspace{-8mm}
    \centering
    \begin{subfigure}{0.45\textwidth}
        \includegraphics[width=\columnwidth]{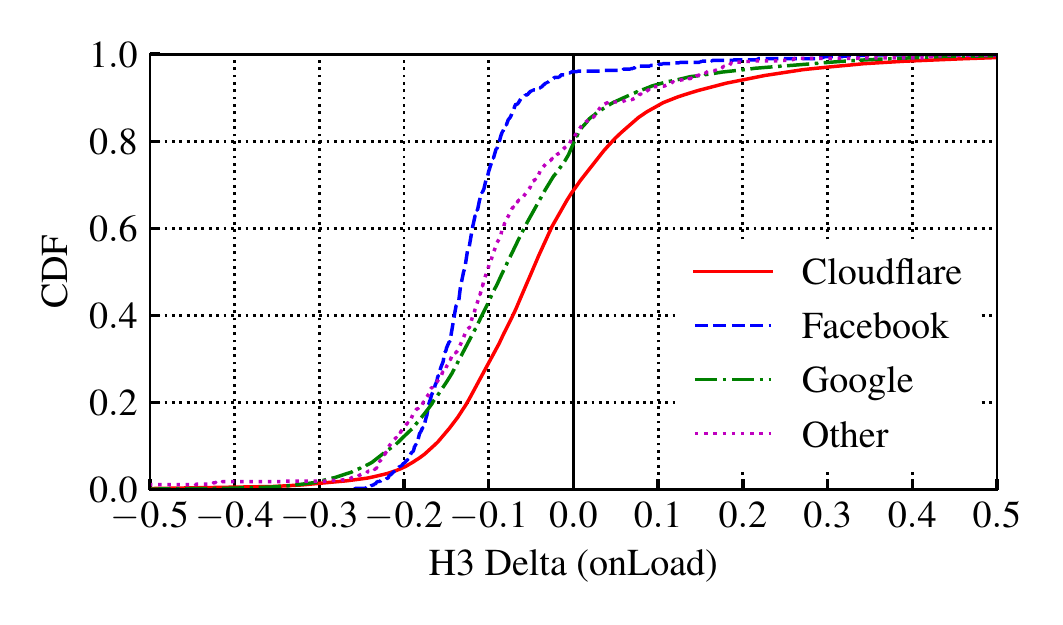}
        \vspace{-5mm}
        \caption{Latency.}
        \label{fig:provider_latency}
    \end{subfigure}
    \begin{subfigure}{0.45\textwidth}
        \includegraphics[width=\columnwidth]{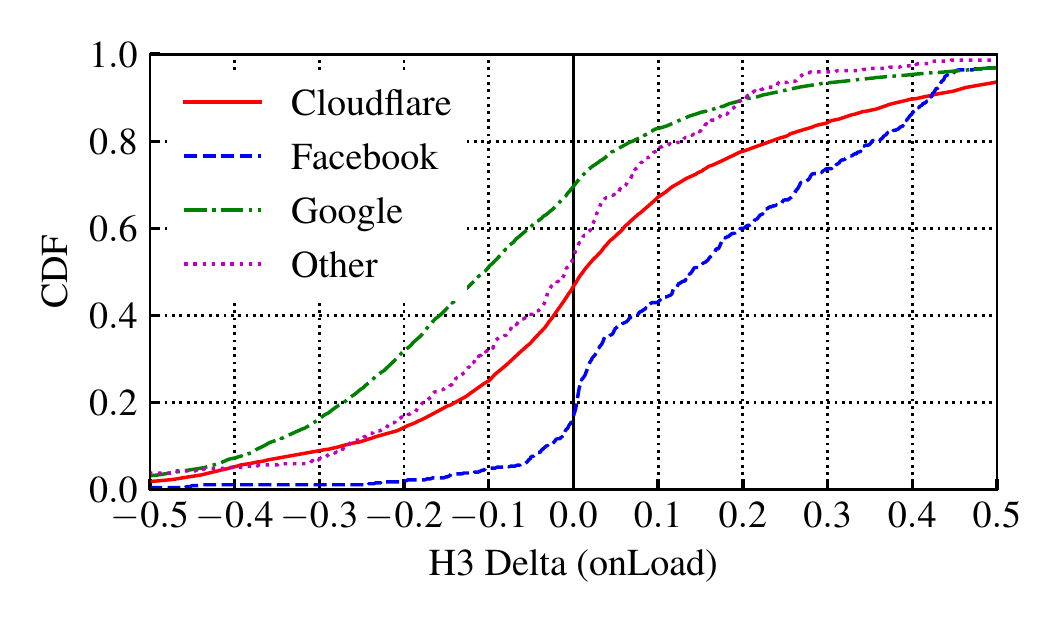}
        \vspace{-5mm}
        \caption{Bandwidth.}
        \label{fig:provider_bw}
    \end{subfigure}
    \caption{onLoad \emph{H3 Delta} by website provider for scenarios with extra-latency and bandwidth limit.}
    \label{fig:provider}
\end{figure}

Figure~\ref{fig:provider} shows the distribution of \emph{H3 Delta} for onLoad, separately by provider. We focus on scenarios with $200$ ms extra latency and $1$ Mbit/s bandwidth limit. We observe that the \emph{H3 Delta} largely differs by provider. Focusing on latency (Figure~\ref{fig:provider_latency}), Facebook websites show the highest performance gain (\emph{H3 Delta} $-0.13$ in median), represented in the figure with the blue dashed line. Moreover, $95$\% of websites are loaded faster with HTTP/3 than with HTTP/2. Cloudflare (red solid line) shows the smallest benefits, with only $72$\% of websites loading faster. Google and the remaining websites sit in the middle. Similar considerations hold for SpeedIndex, not shown here for brevity.

With limited bandwidth (Figure~\ref{fig:provider_bw}), we observe a very different situation. Here, Facebook has in general worst performance with HTTP/3 with $91$\% of its websites loading faster with HTTP/2. Conversely, Google (green dashed line) shows the best figures, with median \emph{H3 Delta} $-0.14$ and $79$\% of websites loading faster with HTTP/3. Cloudflare and the remaining websites exhibit no clear trend with roughly half of the websites loading faster with HTTP/3. 

\begin{figure}[t]
    \centering
    \vspace{-8mm}
    \begin{subfigure}{0.48\textwidth}
        \includegraphics[width=\columnwidth]{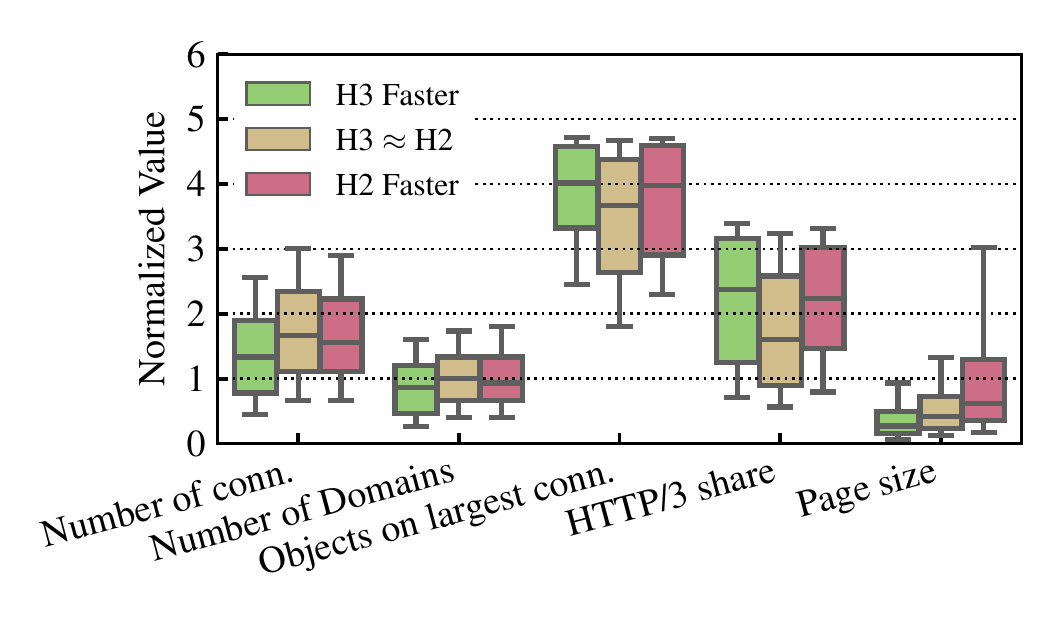}
        \vspace{-5mm}
        \caption{Latency.}
        \label{fig:features_latency}
    \end{subfigure}
    \begin{subfigure}{0.48\textwidth}
        \includegraphics[width=\columnwidth]{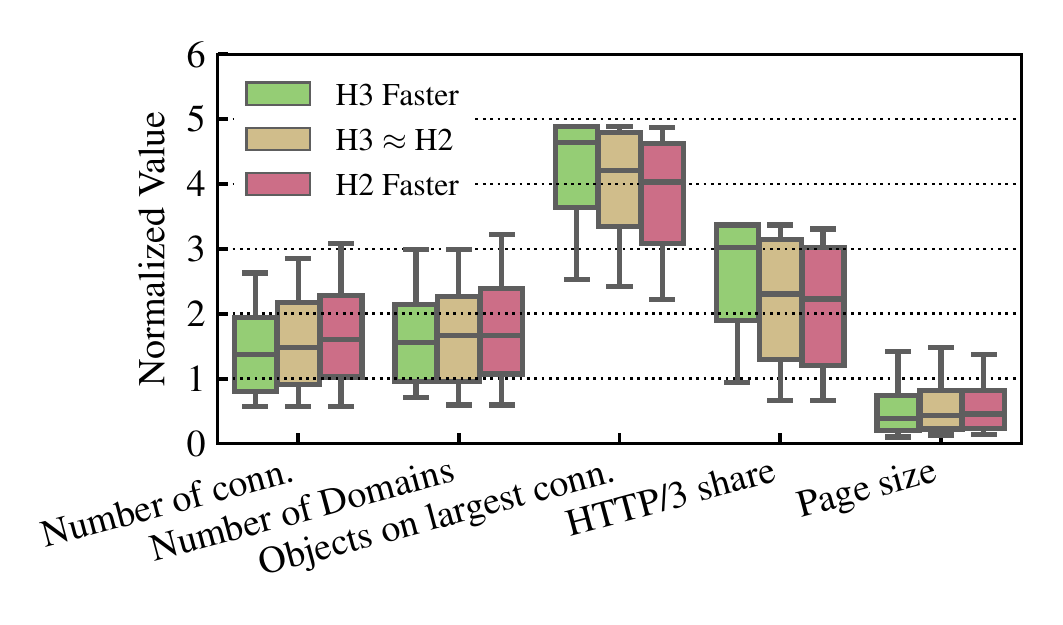}
        \vspace{-5mm}
        \caption{Bandwidth.}
        \label{fig:features_bw}
    \end{subfigure}
    \caption{Visit characteristics vs. \emph{H3 Delta} class (normalized values).}
    \label{fig:features}
\end{figure}

\subsection{Page characteristics}
\label{sec:website}

Now we investigate page characteristics and possible correlations to performance when using HTTP/3. To this end, we compute various metrics describing the web page load process and contrast them to understand whether they show correlations to the \emph{H3 Delta}. For each visit to the $14\,707$ websites in our dataset, in addition to the \emph{H3 Delta}, we compute the following metrics:
\begin{itemize}
    \item {\bf Number of connections} issued by the browser to load the website when using HTTP/3.
    \item {\bf Number of domains} contacted while loading the web page, thus including third-party domains).
    \item {\bf Share of objects on the largest connection}, which measures the share of objects carried over the connection where most objects have been requested. Remind that the best practices of HTTP/3 recommend avoiding domain sharding to increase performance.
    \item {\bf Share of objects served on HTTP/3}, which is used to investigate possible correlations of the fraction of objects served over HTTP/3 (shown in Figure~\ref{fig:share_objects}) and the \emph{H3 Delta} metric. 
    \item {\bf Page Size}, to breakdown performance for small and large web pages. 
\end{itemize}

In Figure~\ref{fig:features}, we compare the distribution of the aforementioned metrics, grouping websites based on classes defined by the onLoad \emph{H3 Delta}:
\begin{itemize}
    \item \textbf{H3 Faster}: websites loading faster with HTTP/3, i.e., onLoad \emph{H3 Delta} $< -0.1$. 
    \item \textbf{H3 $\approx$ H2}: websites having a similar loading time in HTTP/2 and HTTP/3, i.e., onLoad \emph{H3 Delta}~$\in~[-0.1,0.1]$.
    \item \textbf{H2 Faster}: websites loading faster with HTTP/2, i.e., onLoad \emph{H3 Delta}~$> 0.1$.
\end{itemize}

In the figure, boxes with different colors represent these three classes. The $y$-axis represents the metrics normalized by scaling to unit variance to ease the visualization and the comparison. Again, we study the scenarios with extra latency (top row) and limited bandwidth (bottom row) since they provide the most interesting insights. In case of extra latency, H3 Faster websites are $32$\%, H2 Faster $11$\% and H3 $\approx$ H2 are $57$\%. With limited bandwidth, they are $38$\%, $25$\% and $37$\%, respectively.

We first focus on the left-most box group of Figure~\ref{fig:features}, showing the (normalized) number of connections the browser issued to load the web page. Green boxes hint that websites issuing fewer connections (smaller metric values) are faster with HTTP/3 than in HTTP/2. This is true in both scenario, i.e., low latency and poor bandwidth. Similar considerations hold if we focus on the second box group, representing the number of contacted domains. Indeed, we notice that the number of connections per website and the number of contacted domains are $0.91$-correlated (Pearson correlation). The third box group offers a similar perspective, measuring how web page objects are split on multiple connections/domains. The websites benefiting the most from HTTP/3 are those which tend to mass objects on a single connection -- see the highest position of the green boxes, meaning more objects are on a single connection. This is very clear with limited bandwidth (Figure~\ref{fig:features_bw}), rather than with high latency (Figure~\ref{fig:features_latency}).

Serving most objects with HTTP/3 (rather than with HTTP/2) has a positive impact too, as we notice from the fourth box group in Figure~\ref{fig:features}. Again, this is evident especially with bandwidth limit (Figure~\ref{fig:features_bw}), while with extra latency (Figure~\ref{fig:features_latency}) it is hard to find a clear trend. Finally, interesting is the case of the web page size (last box group). In scenarios with high latency, the websites benefiting from HTTP/3 are small, while large ones typically perform better with HTTP/2. Conversely, when bandwidth is scarce, even if moderately, website loading faster with HTTP/3 are the large ones.

In summary, websites taking benefits from HTTP/3 are those limiting the number of connections and third-party domains, and fully adopting HTTP/3 on all web page objects. Page size has diverse implications depending on the network conditions. These considerations hold in scenarios with high latency or limited bandwidth, while we do not observe any clear trend in case of optimal network conditions or high packet loss, where metric distributions mostly overlap.

\section{Discussion and future work}
\label{sec:limitations}

We dissected the performance of HTTP/3 under diverse network conditions, showing the impact of network latency and bandwidth across websites. However, we run measurements using only Google Chrome, as we are not aware of other browsers that can be instrumented to use HTTP/3 since the first connection -- i.e., without the need to observe the \texttt{Alt-Svc} header previously. Moreover, we always visit websites with a fresh browser profile with empty cache and no pre-existing connections. Clearly, this setup limits the scope of our study as we cannot measure how HTTP/3 affects performance on subsequent visits or with a warm HTTP cache, as it will be the case for future users supporting HTTP/3. 

We limited ourselves to a subset of the websites adopting HTTP/3. Indeed, we included only a fraction of websites hosted on the CloudFlare CDN, as its HTTP/3 support is partially disabled at the time of writing. Our measurements will need to be run continuously to observe how the web ecosystem will react to HTTP/3 in the near future, adopting the protocol and modifying the web page structure to optimize performance, in particular concerning domain sharding.

Finally, HTTP/3 and QUIC are not yet definitive IETF standards. Although recent modifications to the IETF draft only concerned minor protocol features, it will be fundamental to provide a similar analysis once the final standards are approved. Similarly, we did not explore how different endpoint configurations affect HTTP/3 performance -- e.g., the interactions of HTTP/3 and congestion control settings. Moreover, whereas we covered different scenarios, it is widely known that network emulation is hard. Similar measurement studies with actual end-users are still needed to confirm our findings. 

\section{Conclusions}
\label{sec:conclusion}

We provided the first study on HTTP/3 adoption and performance, quantifying the performance benefits of HTTP/3 in several network scenarios. We testified how some of the Internet leading companies started deploying HTTP/3 during 2020, although most of the early adopters still host the majority of objects on HTTP/2 third-party servers. With a large-scale measurement campaign, we studied the performance of HTTP/3 under different network conditions, targeting thousands of websites. We found performance benefits emerging in scenarios with high latency or poor bandwidth. In the case of high packet loss, HTTP/3 and HTTP/2 perform roughly the same. We found large performance diversity depending on the infrastructure hosting the website. In general, we observed that websites taking benefits from HTTP/3 are those loading objects from a limited set of third-party domains, thus limiting the number of issued connections and avoiding loading content using legacy protocols.

\bibliographystyle{ieeetr}
\bibliography{h3}

\end{document}